\documentclass[twocolumn,letter]{jpsj2}
\usepackage{bm,times,mathptmx}

\title{
Charge-Density-Wave Formation in the Doped Two-Leg Extended Hubbard Ladder 
}

\author{Masahisa {\sc Tsuchiizu} and Yoshikazu {\sc Suzumura}}

\inst{
Department of Physics, Nagoya University, Nagoya 464-8602, Japan
}

\abst{
We investigate electronic properties 
  of the doped two-leg Hubbard ladder 
  with both the onsite and the nearest-neighbor Coulomb repulsions,
  by using the the weak-coupling renormalization-group method.
It is shown that,
for strong nearest-neighbor repulsions,
the charge-density-wave state coexisting with the 
$p$-density-wave state becomes dominant fluctuation 
 where spins form intrachain singlets.
By increasing doping rate,
we have also shown that the effects of the nearest-neighbor repulsions  
   are reduced and 
the system exhibits a quantum phase transition into
   the $d$-wave-like (or rung-singlet) superconducting state.
We derive the effective fermion theory which describes the critical properties 
  of the transition point
  with the gapless excitation of  magnon. 
The phase diagram of the two-leg ladder compound,
Sr$_{14-x}$Ca$_x$Cu$_{24}$O$_{41}$,  is  discussed.
}

\kword{doped Hubbard ladder, intersite Coulomb repulsion,
spin gap, charge density wave, superconductivity}

\begin{document}
\maketitle

Electronic properties on ladder systems have been studied intensively
 both theoretically and  experimentally, since
  the superconducting (SC) state was discovered in
  the self-doped two-leg ladder material Sr$_{14-x}$Ca$_x$Cu$_{24}$O$_{41}$ 
  with  $x \gtrsim 12$ under pressure over 3 GPa.\cite{Uehara,Nagata}    
The substitution of Ca for Sr changes effectively the
  hole-doping rate in the ladder Cu sites, 
  where the  rate
  varies monotonically from 0.07 to 0.25 with increasing $x$ from 0 to
   12. \cite{Osafune}
A characteristic feature is the presence of a gap in 
  magnetic excitations at temperature much higher than 
  the SC transition temperature.
  \cite{Piskunov,Fujiwara}
Besides the SC state,
recent experimental studies have focused on 
  the charge dynamics in the slightly doped materials and 
  verified collective modes from 
  the sliding of the charge-density-wave (CDW) developed on ladder sites.
   \cite{Kitano,Blumberg,Gorshunov,Vuletic} 
  A global phase diagram is obtained on the plane of  $x$ and temperature
   showing that the hole doping suppresses the CDW state 
   followed by  the insulating state without the CDW order, 
   and then the high doping leads to the SC state under pressure. 
\cite{Vuletic}
Quite recently the CDW collective modes are also suggested in the 
  highly doped material Sr$_2$Ca$_{12}$Cu$_{24}$O$_{41}$.\cite{Gozar} 
Therefore it is of particular interest to investigate 
  the competition between the SC state and the  CDW state in doped
   ladder systems.
  
From a theoretical point of view,
the origin of the spin gap in the ladder compounds
  seems to be explained successfully for
 both the undoped\cite{Dagotto,Schulz} and doped\cite{Schulz} cases,
  and it is known that
   the $d$-wave-like SC (SC$d$)
   state appears in doped ladder systems.  \cite{Schulz}
In addition, the charge-ordered state is also suggested 
 when intersite interactions are included. \cite{Vojta}
Further the competition 
  between the SC$d$ state and the charge-ordered or charge-density-wave (CDW) 
  state has been examined.\cite{Schulz,Vojta,Fradkin2002}
 However, critical behavior is not yet fully understood.
In the present paper,   
  the possible scenario of the instability of the CDW state and the
 competition between the CDW state and the  
  SC state are proposed 
in the doped two-leg ladder of
  the extended Hubbard model (EHM)
  with nearest-neighbor repulsive interactions  
   by extending the previous analytical calculations 
   \cite{Tsuchiizu2002b,Fradkin2002} 
  to analyze the critical behavior in more detail.

We consider the two-leg EHM given by $H=H_0+H_{\mathrm{int}}$.
 The first term describes the hopping energies along and between legs:
\begin{eqnarray}
H_0
 \!\!\!\! &=& \!\!\!\! 
 - t_\parallel  \sum_{j,\sigma,l}
   (c_{j,l,\sigma}^\dagger \, c_{j+1,l,\sigma}^{}+\mathrm{H.c.})
\nonumber \\
&& {} \!\!\!\! - t_\perp \sum_{j,\sigma}
   (c_{j,1,\sigma}^\dagger \, c_{j,2,\sigma}^{}+\mathrm{H.c.})
,
\label{eq:H0}
\end{eqnarray}
where $c_{j,l,\sigma}$ annihilates an electron of spin 
  $\sigma(=\uparrow,\downarrow)$ on rung $j$ and leg $l(=1,2)$.
The Hamiltonian $H_{\mathrm{int}}$ denotes interactions between electrons:
\begin{equation}
H_{\mathrm{int}}
=
U \sum_{j,l} n_{j,l,\uparrow} \, n_{j,l,\downarrow}
+V_\parallel \sum_{j,l} n_{j,l} \, n_{j+1,l}
+V_\perp \sum_{j} n_{j,1} \, n_{j,2},
\end{equation}
  where $U$ represents on-site repulsion and 
  $V_\parallel$ ($V_\perp$) represents  intrachain (interchain) 
  nearest-neighbor repulsion with 
   $n_{j,l,\sigma}=c_{j,l,\sigma}^\dagger \, c_{j,l,\sigma}^{}$
   and $n_{j,l}=n_{j,l,\uparrow}+n_{j,l,\downarrow}$.
The $H_0$ term is diagonalized by using the Fourier transform of  
  $c_{\sigma}(\bm{k})$   
  where $\bm{k}=(k_\parallel,k_\perp)$ with $k_\perp = 0$ or $\pi$.
The energy dispersion is given by
 $\varepsilon(\bm{k}) = -2 t_\parallel \cos k_\parallel
       - t_\perp \cos k_\perp$.
Here we consider the case with finite hole doping $\delta$   
  satisfying $t_\perp< 2t_\parallel \cos^2 \frac{\pi}{2}\delta$,
  in which both the bonding ($k_\perp=0$) and the antibonding
  ($k_\perp=\pi$) energy bands are partially filled and
  the Fermi points are located at 
 $   k_{F,0}=\frac{\pi}{2}(1-\delta)+\lambda$ and at
 $  k_{F,\pi}=\frac{\pi}{2}(1-\delta)-\lambda$ with
  $\lambda \equiv \sin^{-1} 
   \left[t_\perp/\left(2t_\parallel\cos \frac{\pi}{2}\delta \right)\right]$.
We examine the case of small $\delta$ by 
 neglecting the differences in the Fermi velocities of  the
bonding/antibonding band, i.e., $v_{F,0}=v_{F,\pi} (\equiv v_F)$.

Following the standard weak-coupling approach ($g$-ology),
the linearized kinetic energy is given by
$ H_0 = \sum_{\bm{k},p,\sigma}
v_F (pk_\parallel-k_{F,k_\perp})  \,
  c_{p,\sigma}^\dagger (\bm{k}) \, c^{}_{p,\sigma} (\bm{k})$,
   where the index $p=+/-$ denotes the right-/left-moving electron.
By introducing field operators  by
  $\psi_{p,\sigma,\zeta}(x)=
   L^{-1/2}\sum_{k_\parallel} e^{ik_\parallel x}
   c_{p,\sigma}(k_\parallel,k_\perp)$ 
  with $\zeta=+(-)$ for $k_\perp=0(\pi)$
  and $L$ being the system size,
   the interactions near the Fermi points are rewritten as 
  $H_{\mathrm{int}}= (1/4)
  \int dx  
   \sum_{p,\sigma}{\sum_{\zeta_i=\pm}}'
  \mathcal{H}_{\mathrm{int}}$, where  $\mathcal{H}_{\mathrm{int}}$ is given by 
\begin{eqnarray}
 & &  \!\!\!\!
      g_{1(2)\parallel}^{\epsilon\bar\epsilon} \,
      \psi_{p,\sigma,\zeta_1}^\dagger \,
      \psi_{-p,\sigma,\zeta_2}^\dagger \,
      \psi_{+(-)p,\sigma,\zeta_4}^{} \,
      \psi_{-(+)p,\sigma,\zeta_3}^{}
\nonumber \\
  &+& \!\!\!\!
      g_{1(2)\perp}^{\epsilon\bar\epsilon} \,
      \psi_{p,\sigma,\zeta_1}^\dagger \,
      \psi_{-p,\bar\sigma,\zeta_2}^\dagger \,
      \psi_{+(-)p,\bar\sigma,\zeta_4}^{} \,
      \psi_{-(+)p,\sigma,\zeta_3}^{} ,
\quad
\label{eq:Hint_g-ology}
\end{eqnarray} 
   and  $\bar\sigma=\uparrow(\downarrow)$ for 
   $\sigma=\downarrow(\uparrow)$, 
  $\epsilon=\zeta_1\zeta_3$ and $\bar\epsilon=\zeta_1\zeta_2$.
The summation of the band index $\zeta_i$ ($i=1,\ldots,4$) is taken
   under the condition $\zeta_1 \zeta_2 \zeta_3 \zeta_4 = +1$. 
The coupling constants $g_{i\parallel}^{\epsilon\bar\epsilon}$
 and $g_{i\perp}^{\epsilon\bar\epsilon}$ 
 with $i=1 (2)$ corresponding  to the backward (forward) scattering
 are given by 
$g_{i\parallel}^{\epsilon\bar\epsilon}=
  (l_\epsilon V_\perp+ m_{i,\epsilon} V_\parallel)$ and 
$g_{i\perp}^{\epsilon\bar\epsilon}
= (U + l_\epsilon  V_\perp
  + m_{i,\epsilon} V_\parallel)$
   where $l_\pm = \pm 1$, $m_{1,+}=-2\cos\pi\delta \, \cos 2\lambda$,
   $m_{1,-} =-2\cos\pi\delta$, $m_{2,+}=+2$, and $m_{2,-}=+2\cos 2\lambda$.
We neglect the umklapp scattering processes which become irrelevant for
   finite doping case  and also neglect forward scattering processes which 
   do not yield qualitative changes in the system. \cite{Emery}  

As possible states, 
 we consider the singlet $d$-wave superconducting (SC$d$) state, 
  the CDW state, and the $p$-density-wave (PDW) state.  \cite{Tsuchiizu2002b}
  The PDW state corresponds to the spin-Peierls state
   in the limit of $\delta\to 0$.
 The order parameter of the SC$d$ state is given by
$O_{\mathrm{SC}d} = N^{-1} \sum_{j}
   (c_{j,1,\uparrow} \, c_{j,2,\downarrow}
    - c_{j,1,\downarrow} \, c_{j,2,\uparrow})$,
 while those of the density waves are 
$ O_A = N^{-1}\sum_{\bm{k},\sigma} 
  f_A (\bm{k}) \, c_\sigma^\dagger (\bm{k}) \, c_\sigma^{} (\bm{k+Q})$,
  with  $\bm{Q}=\bigl(\pi(1-\delta),\pi\bigr)$,  
  $f_{\mathrm{CDW}}=1$ and
  $f_{\mathrm{PDW}}=\sin k_\parallel$. 
These operators are rewritten in terms of bosonic phase fields by  
  applying the Abelian bosonization method\cite{Emery,Gogolin_book}.
The field operators of the right- and left-moving electrons are written as
$\psi_{p,\sigma,\zeta}(x) =
\eta_{\sigma,\zeta}(2\pi a)^{-1/2} \,
\exp [ ipk_{F,k_\perp} x 
 + i p\, \varphi _{p,s,\zeta}(x) ] $
   where $s=+$ for $\sigma=\uparrow$ and $s=-$ for
   $\sigma=\downarrow$ and the fields satisfy 
   the commutation relations:
   $[\varphi_{p,s,\zeta}(x),\varphi_{p,s',\zeta'}(x')]
    = ip\pi \, \mathrm{sgn}(x-x') \, 
    \delta_{s,s'}\,\delta_{\zeta,\zeta'}$
   and
   $[\varphi_{+,s,\zeta},\varphi_{-,s',\zeta'}]
    = i\pi \,\delta_{s,s'}\,\delta_{\zeta,\zeta'}$.
The Klein factors $\eta_{\sigma,\zeta}$ 
   are introduced in order to retain the correct anticommutation
   relations.\cite{Tsuchiizu2002b}
   For calculating physical quantities, the field 
 $\varphi _{p,s,\zeta} $ is replaced by new  bosonic fields:
$\phi_{\nu r}=(\phi_{\nu r}^+ + \phi_{\nu r}^-)$ and 
  $\theta_{\nu r}=(\phi_{\nu r}^+ - \phi_{\nu r}^-)$ 
where 
$\varphi_{p,s,\zeta}  =
 (\phi_{\rho +}^p  + \zeta  \phi_{\rho -}^p
  + s \phi_{\sigma +}^p    + s \zeta \phi_{\sigma -}^p)$ with
     $p=\pm$, $s=\pm$, and $\zeta=\pm$.
The phase fields $\phi_{\rho\pm}$ and $\phi_{\sigma\pm}$ represent charge 
   and spin fluctuations, respectively and the suffices $\pm$ refers to
   the even and odd sectors. They satisfy
   $[ \phi_{\nu r}(x), \theta_{\nu' r'}(x') ] =
   -i \pi \Theta(-x+x')\delta_{r,r'}$
   with $\Theta(x)$ being the Heaviside step function.
In terms of $\phi_{\nu r}$ and  $\theta_{\nu r}$,
   the order parameters $O=\int dx \mathcal{O}$ are given by
\begin{subequations}
\begin{eqnarray}
&&
\mathcal{O}_{\mathrm{SC}d}
\propto 
   e^{i \theta_{\rho+}}
  \cos \theta_{\rho-} \,
  \cos \phi_{\sigma+} \,
  \cos \phi_{\sigma-}
\nonumber \\ && \qquad\qquad
{}  - i \,
  e^{i \theta_{\rho+}}
  \sin \theta_{\rho-} \,
  \sin \phi_{\sigma+} \,
  \sin \phi_{\sigma-} 
,
\\ 
&&
\mathcal{O}_{\mathrm{CDW}}
\propto
  \cos \phi_{\rho+} \,
  \sin \theta_{\rho-} \,
  \cos \phi_{\sigma+} \,
  \cos \theta_{\sigma-} 
\nonumber  , \\ 
&& \qquad\qquad
{} -
  \sin \phi_{\rho+} \,
  \cos \theta_{\rho-} \,
  \sin \phi_{\sigma+} \,
  \sin \theta_{\sigma-} ,
\label{eq:order_CDW}
\\
&&
\mathcal{O}_{\mathrm{PDW}}
\propto
  \cos \phi_{\rho+} \,
  \cos \theta_{\rho-} \,
  \sin \phi_{\sigma+} \,
  \sin \theta_{\sigma-} ,
\nonumber \\ 
&& \qquad\qquad
{} +
  \sin \phi_{\rho+} \,
  \sin \theta_{\rho-} \,
  \cos \phi_{\sigma+} \,
  \cos \theta_{\sigma-} .
\qquad\quad
\end{eqnarray}%
\label{order-parameters}%
\end{subequations}

We can also rewrite 
  the Hamiltonian in terms of the bosonic phase variables.  
In Eq.\ (\ref{eq:Hint_g-ology}),  
  the phase field $\phi_{\rho-}$ appears
  in the form $\cos (2\phi_{\rho-}+4\lambda x)$.
Since we can safely assume that  $t_\perp$ is a relevant perturbation
  for $t_\perp$  being  not so small \cite{Tsuchiizu1999,Tsuchiizu2001},
  the term with $\cos (2\phi_{\rho-}+4\lambda x)$ 
  would become irrelevant, and thus we discard it in the following.
We also neglect the  $\cos 2\phi_{\sigma-} \, \cos 2\theta_{\sigma-}$ term 
 since its scaling dimension is larger than 2. 
Then our Hamiltonian reduces to $H=\int dx \mathcal{H}$ with
\begin{eqnarray}
{\cal H}
\!\!\!\! &=& \!\!\!\! 
\frac{v_F}{\pi}  \sum_{r=\pm}
\Bigl[
\sum_{p=\pm}  \left(  \partial_x \phi_{\rho r}^p \right)^2
+ \frac{g_{\rho r}}{\pi v_F} 
   (\partial_x \phi_{\rho r}^+ )
   (\partial_x \phi_{\rho r}^- )
\Bigr]
\nonumber \\ && \!\!\!\! {}
+
\frac{v_F}{\pi}
 \sum_{r=\pm}
\Bigl[
\sum_{p=\pm}  \left(  \partial_x \phi_{\sigma r}^p \right)^2
- \frac{g_{\sigma r}}{\pi v_F} 
   \left(\partial_x \phi_{\sigma r}^+ \right)
   \left(\partial_x \phi_{\sigma r}^- \right)
\Bigr]
\nonumber \\
&&\!\!\!\! {} 
+ \frac{1}{2\pi^2 a^2}
(
  g_{\overline{c-},s+} \,
    \cos 2 \phi_{\sigma+}
+   g_{\overline{c-},s-}\,
    \cos 2 \phi_{\sigma-} 
  \nonumber \\
&& {} \qquad\qquad
+   g_{\overline{c-},\overline{s-}}\,
    \cos 2 \theta_{\sigma-}
) \, \cos 2 \theta_{\rho-} 
\nonumber \\
&&\!\!\!\! {} 
+ \frac{1}{2\pi^2 a^2}
(
   g_{s+, s-}\,
    \cos 2 \phi_{\sigma+} \,
    \cos 2 \phi_{\sigma-}   
  \nonumber \\
&& {} \qquad\qquad
+   g_{s+,\overline{s-}}\,
    \cos 2 \phi_{\sigma+}\,
    \cos 2 \theta_{\sigma-}  
),
\label{eq:Hboson}
\end{eqnarray}
where the coupling constants of the harmonic terms are given by  
$g_{\rho (\sigma) r} = \sum_{\epsilon=\pm} f_{\rho(\sigma)r}^\epsilon
   ( g_{2\parallel}^{+\epsilon} + (-) g_{2\perp}^{+\epsilon}
    -g_{1\parallel}^{\epsilon\epsilon})/2 $  
  with $r=\pm$,  
   $f_{\rho+}^\epsilon=1$, $f_{\rho-}^\epsilon=\epsilon$, 
       $f_{\sigma+}^\epsilon=-1$ and $f_{\sigma-}^\epsilon=-\epsilon$. 
 The coupling constants of the nonlinear terms are 
  $g_{\overline{c-},s+}\equiv - g_{1\perp}^{-+}$,
  $g_{\overline{c-},s-}\equiv - g_{2\perp}^{-+}$,
  $g_{\overline{c-},\overline{s-}}\equiv
     (g_{2\parallel}^{-+} - g_{1\parallel}^{-+})$, 
  $g_{s+,s-} \equiv  g_{1\perp}^{++}$,
  $g_{s+,\overline{s-}} \equiv  g_{1\perp}^{--}$. 
These nine coupling constants are not independent 
since  the global spin-rotation SU(2) symmetry leads to
\cite{Tsuchiizu2002b} 
$(g_{\sigma+}+g_{\sigma-} - g_{s+,s-}) = 
 (g_{\sigma+}-g_{\sigma-} - g_{s+,\overline{s-}}) = 
 (g_{\overline{c-},s+} - g_{\overline{c-},s-}
-g_{\overline{c-},\overline{s-}}) = 0$.
We choose following six  coupling constants: 
\begin{subequations}
\begin{eqnarray}
&& \hspace*{-.5cm}
g_{\overline{c-},st}
 = 
+U-V_\perp -2 V_\parallel \cos \pi\delta ,
, \\
&& \hspace*{-.5cm}
g_{\overline{c-},ss}
= {} +U-V_\perp + 2V_\parallel(\cos \pi \delta + 2\cos 2\lambda), \qquad
\\
&& \hspace*{-.5cm}
g_{\rho+} =
+U+2V_\perp +V_\parallel [4+\cos\pi\delta(1+\cos2\lambda)],
\\
&& \hspace*{-.5cm}
g_{\rho-} =
-V_\perp - V_\parallel \cos \pi\delta (1- \cos 2\lambda),
\\
&& \hspace*{-.5cm}
g_{\sigma+} =
+U - V_\parallel \cos\pi\delta (1+\cos 2\lambda ),
\\
&& \hspace*{-.5cm}
g_{\sigma-}  =
+V_\perp + V_\parallel \cos\pi\delta (1-\cos 2\lambda ).
\end{eqnarray}%
\label{eq:g}%
\end{subequations}
where 
$g_{\overline{c-},st}
\equiv -g_{\overline{c-},s+}$ and 
$ g_{\overline{c-},ss}
\equiv  (-g_{\overline{c-},s-}+g_{\overline{c-},\overline{s-}})$.
The present model and the above treatment are quite similar to those
  in Ref.\ \citen{Fradkin2002}.
However, the application of the  renormalization-group (RG) method 
to Eq.\ (\ref{eq:Hboson})
  is complicated to estimate excitation gaps of spin modes properly.
Therefore, 
we fermionize the spin part of Eq.\ (\ref{eq:Hboson})  \cite{Tsuchiizu2002b}
 by introducing
   spinless fermion fields 
$\psi_{\pm,r}(x) = 
  \eta_r (2\pi a)^{-1/2}
  \, \exp\left[ \pm i \,2\phi_{\sigma r}^{\pm}(x)\right]$
   where  $r=\pm$ and $\{\eta_r,\eta_{r'}\}=2\delta_{r,r'}$.
By using the SU(2) constraints and 
 the Majorana fermions $\xi^n$ ($n=1\sim 4$), 
the equation (\ref{eq:Hboson}) is rewritten as
\begin{eqnarray}
\mathcal{H}
\hspace*{-.3cm}&=&\hspace*{-.3cm} {}
\frac{v_F}{\pi} \sum_{r}
\Bigl[
   \sum_p \left(\partial \phi_{\rho r}^p \right)^2 
 + \frac{g_{\rho r}}{\pi v_F} 
   (\partial_x \phi_{\rho r}^+ )
   (\partial_x \phi_{\rho r}^- )
\Bigr]
\nonumber \\ && {}
-i\frac{v_F}{2} 
\left(
  \bm{\xi}_+ \cdot \partial_x \bm{\xi}_+
- \bm{\xi}_- \cdot \partial_x \bm{\xi}_-
\right)
-\frac{g_{\sigma+}}{2} \, 
\left(
 \bm{\xi}_+ \cdot \bm{\xi}_-
\right)^2
\nonumber \\ && {}
-i\frac{v_F}{2}
\left(
 \xi_+^4 \, \partial_x \xi_+^4
- \xi_-^4 \, \partial_x \xi_-^4
\right)
-g_{\sigma-} \,
 \left( \bm{\xi}_+ \cdot \bm{\xi}_- \right)
 \, \xi_+^4 \, \xi_-^4
\nonumber \\ && {}
- \frac{i}{2\pi a}
\left(
g_{\overline{c-},st} \,
  \bm{\xi}_+ \cdot \bm{\xi}_-
+
g_{\overline{c-},ss} \,
  \xi_+^4 \cdot \xi_-^4
\right)  \,
\cos 2\theta_{\rho-} ,
\label{eq:Heff}
\end{eqnarray}
where 
$\psi_{p,+}= (\xi_{p}^1+i\xi_{p}^2 )/\sqrt{2}$, 
$\psi_{p,-} = (\xi_{p}^4+i\xi_{p}^3 )/\sqrt{2}$,
and $\bm{\xi}_p=(\xi_p^1,\xi_p^2,\xi_p^3)$.
Thus the effective theory for the spin sector becomes
   O(3)$\times$Z$_2$ symmetric, as seen in the 
  isotropic Heisenberg \cite{Gogolin_book} and half-filled Hubbard ladder.
  \cite{Tsuchiizu2002b}

We investigate the low-energy behavior by using the perturbative
  RG method with  the lattice constant $a\to a e^{dl}$. 
 Following six scaling equations are obtained:
\begin{subequations}
\begin{eqnarray}    
&& \hspace*{-1.2cm}
\frac{d}{dl} G_{\rho-} =
 - \frac{3}{4} G_{\overline{c-},st}^2
 - \frac{1}{4} G_{\overline{c-},ss}^2
,
\\ 
&& \hspace*{-1.2cm}
\frac{d}{dl} G_{\sigma+} =
 - G_{\sigma+}^2
 - G_{\sigma-}^2
 -\frac{1}{2} G_{\overline{c-},st}^2
,
\\ 
&&\hspace*{-1.2cm}
\frac{d}{dl} G_{\sigma-} =
 - 2 G_{\sigma+} \, G_{\sigma-}
 - \frac{1}{2} G_{\overline{c-},st} \, G_{\overline{c-}ss}
,
\\
&&\hspace*{-1.2cm}
\frac{d}{dl} G_{\overline{c-},st} =
  -  G_{\rho-} \, G_{\overline{c-},st}
  - 2 G_{\sigma+} \, G_{\overline{c-},st}
  - G_{\sigma-} \, G_{\overline{c-},ss} ,
\\
&&\hspace*{-1.2cm}
\frac{d}{dl} G_{\overline{c-},ss} =
  - G_{\rho-} \, G_{\overline{c-},ss}
  -3  G_{\sigma-} \, G_{\overline{c-},st}
,
\end{eqnarray}
\end{subequations}
and $dG_{\rho+}/dl =0$ where $G(0)=g/(2\pi v_F)$.
We noted that these RG equations can be also derived directly from 
  Eq.\ (\ref{eq:Hboson}).
We analyze the RG equations numerically for 
 $ U>0$, $ V_{\parallel}>0 $ and  
 $ V_{\perp} > 0 $. 
For small  $V_{\perp}/U$ and $V_{\parallel}/U$, 
  the limiting behavior of RG equations is given by  
  $(G_{\rho-}^*,G_{\sigma+}^*,G_{\sigma-}^*,
    G_{\overline{c-},st}^*,G_{\overline{c-},ss}^*)
   =(-,-,-,+,+)$
 which corresponds  to 
  $(g_{\overline{c-},s+}^*,g_{\overline{c-},s-}^*, 
    g_{\overline{c-},\overline{s-}}^*,g_{s+, s-}^*, g_{s+,\overline{s-}}^*)
   = (-,-,0,-,0)$  in  Eq.\ (\ref{eq:Hboson}). 
The relevant behavior of coupling constants implies that the phases are 
  locked in order to minimize 
   the cosine potential in Eq.\ (\ref{eq:Hboson}). 
The positions of phase locking and the corresponding ground states
  are summarized in Table \ref{table:phase}.
Since the $\theta_{\sigma-}$ field is  conjugate  to
$\phi_{\sigma-}$,
  these two fields cannot be locked at the same time.
From Eq.\ (\ref{order-parameters}),
  the nonvanishing order parameter is $\mathcal{O}_{\mathrm{SC}d}$.
Since the correlation function of the operator $e^{i\theta_{\rho+}}$ exhibits
  power-law behavior, 
  we obtain that the SC$d$ fluctuation becomes quasi-long-range ordered
  (quasi-LRO) in this case.
We note that  the SC$d$ state moves to  
  the $D$-Mott or $D^\prime$-Mott state
  in the  limit of $\delta\to 0$.\cite{Tsuchiizu2002b} 
For large 
$ V_{\perp}/U$  and $ V_{\parallel}/U$, 
the limiting behavior of RG equations is now given by  
  $(G_{\rho-}^*,G_{\sigma+}^*,G_{\sigma-}^*,
    G_{\overline{c-},st}^*,G_{\overline{c-},ss}^*)
   =(-,-,+,-,+)$, corresponding to
  $(g_{\overline{c-},s+}^*,g_{\overline{c-},s-}^*, 
    g_{\overline{c-},\overline{s-}}^*,g_{s+, s-}^*, g_{s+,\overline{s-}}^*)$
  $ = (+,0,+,0,-)$.
In this case,
  the dominant order parameters are given by 
   $\mathcal{O}_{\mathrm{CDW}}$ and 
  $\mathcal{O}_{\mathrm{PDW}}$ 
 both of which lead to the quasi-LRO with the same 
  exponent of the correlation functions.
We call this coexisting state the CDW+PDW state.

In order to analyze the properties near the critical point of the transition
  between the SC$d$ state and the CDW+PDW state,
  we restrict ourselves to the case where the mass of the charge mode
  ($\rho-$)  is larger than those of the spin modes ($\sigma\pm$). 
The $\theta_{\rho-}$ field is locked by the cosine
   potential below the mass scale of the charge mode $m_{\rho-}$.
By replacing  $\cos2\theta_{\rho-}$ with its average value 
 $c_{\overline{\rho-}}\equiv
  \langle \cos 2\theta_{\rho-}\rangle$ in Eq.\ (\ref{eq:Hboson}), 
 the effective low-energy Hamiltonian for 
  the spin degrees of freedom is obtained as 
     \cite{Gogolin_book,Tsuchiizu2002b} 
\begin{eqnarray}
\mathcal{H}_\sigma 
\!\!\!\!&=&\!\!\!\! {}
-i\frac{v_F}{2} 
\left(
  \bm{\xi}_+ \cdot \partial_x \bm{\xi}_+
- \bm{\xi}_- \cdot \partial_x \bm{\xi}_-
\right)
 - i m_t^0 \,
  \bm{\xi}_+ \cdot \bm{\xi}_-
\nonumber \\ && \!\!\!\! {}
-i\frac{v_F}{2}
\left(
 \xi_+^4 \, \partial_x \xi_+^4
- \xi_-^4 \, \partial_x \xi_-^4
\right)
- i m_s^0 \,
   \xi_+^4 \, \xi_-^4
\nonumber \\ && \!\!\!\! {}
-\frac{g_{\sigma+}}{2} \, 
\left(
 \bm{\xi}_+ \cdot \bm{\xi}_-
\right)^2
- g_{\sigma-}
 \left( \bm{\xi}_+ \cdot \bm{\xi}_- \right)
 \, \xi_+^4 \, \xi_-^4
, \qquad
\label{eq:Heff_spin}
\end{eqnarray}
where $m_t^0$ and $m_s^0$ represent bare masses of the Majorana triplet
  and singlet sector:
$m_t^0= (c_{\overline{\rho -}}/2\pi a)  
( U-V_\perp -2 V_\parallel \cos \pi\delta )$ and
$m_s^0 = (c_{\overline{\rho -}}/2\pi a)
[U-V_\perp + 2V_\parallel (\cos \pi \delta + 2\cos 2\lambda)]$.
The quantity  $m_t^0$ ($m_s^0$) has physical meanings of the gap
  in the magnon (soliton) excitation in the spin modes of the ladder. 
  \cite{Gogolin_book}
Equation (\ref{eq:Heff_spin}) is further analyzed 
in terms of the following scaling equations for coupling constants: 
\begin{subequations}
\begin{eqnarray}
&&
\frac{dG_{t}}{dl} =
G_{t}-2 G_{t}G_{\sigma+}-G_{s}G_{\sigma-},\\
&&
\frac{dG_{s}}{dl} =
G_{s} -3  G_{t}G_{\sigma-},\\
&&
\frac{dG_{\sigma+}}{dl} =
-G_{\sigma+}^2 - G_{\sigma-}^2 - G_{t}^2,\\
&&
\frac{dG_{\sigma-}}{dl} =
-2G_{\sigma+}G_{\sigma-}-G_{t}G_{s},
\end{eqnarray}%
\label{eq:dG_spin}%
\end{subequations}
where  $G_{t}=m_t^0/v_F$, $G_{s}=m_s^0/ v_F$, and
  $G_{\sigma\pm}=g_{\sigma\pm}/2\pi v_F$.
The couplings $G_s$ and $G_t$ are relevant,
 while $G_{\sigma\pm}$ are marginal.
In Eq.\ (\ref{eq:dG_spin}), the $G_s$ term as a function
  of $l$ increases rapidly  compared with other $G's$ 
  and becomes relevant at $l=l_s$ corresponding to the energy scale  
  of a gap in the Majorana singlet mode $m_s \approx t_\parallel
  e^{-l_s}$, where we stop the calculation of Eq.\
  (\ref{eq:dG_spin}).
The mode remained below the energy scale of $m_s$  
is the  Majorana triplet sector. The effective theory for this mode is 
 given by 
$\mathcal{H}_\sigma^{\mathrm{eff}}  =
-i\frac{1}{2}v_F 
(  \bm{\xi}_+ \cdot \partial_x \bm{\xi}_+
 - \bm{\xi}_- \cdot \partial_x \bm{\xi}_-)
 - i m_t^s \,
  \bm{\xi}_+ \cdot \bm{\xi}_-
-\frac{1}{2}g_{\sigma+}^s \, ( \bm{\xi}_+ \cdot \bm{\xi}_- )^2$
 where $m_t^s=v_F [G_t(l_s)-G_{\sigma-}(l_s)]$ and $g_{\sigma+}^s=2\pi
  v_F G_{\sigma+}(l_s)$.
 Then we solve the RG equations given by 
 $dG_t/dl=G_t-2G_tG_{\sigma+}$ and 
  $dG_{\sigma+}/dl=-2G_{\sigma+}^2-G_t^2$ with the initial conditions 
  $G_t(l_s)=m_t^s/v_F$ and  $G_{\sigma+}(l_s)=g_{\sigma+}^s/2\pi v_F$. 
We easily find that these RG equations have two stable fixed points 
  $(G_t^*,G_{\sigma+}^*)=(+\infty,-\infty)$ and $(-\infty,-\infty)$,
  corresponding to the SC$d$ state and the CDW+PDW state, respectively,
  where the magnitude of the gap in the Majorana triplet sector 
   can be estimated from $ m_t \approx t_\parallel e^{-l_t}
  \mathrm{sgn}(G_t^*)$ 
  where $l_t$ is determined by $|G_t(l_t)|=1$
  (see Table \ref{table:phase}).
There are also 
  two unstable fixed points 
   $(G_t^*,G_{\sigma+}^*)=(0,0)$ and $(0,-\infty)$, corresponding 
   to the  second-order and first-order phase transitions,
  \cite{Tsuchiizu2002b}
  while only the former transition is obtained in the present 
  numerical calculation.
\begin{table}[t]
\caption{
Possible states and the corresponding pattern of phase locking 
 where  $I$ is  an integer and 
  the symbol $*$ indicates an unlocked bosonic phase field.
 The signs  $+$ and $-$ denote those for 
 renormalized  masses $m_t$ and $m_s$ where
  we have assumed $I$ being even number, i.e., $c_{\overline{\rho-}}>0$.
}
\label{table:phase}
\begin{tabular}{l|cccc|cc}
\hline\hline
 State   & $\langle\theta_{\rho-}\rangle$ & 
           $\langle\phi_{\sigma+}\rangle$ & 
           $\langle\phi_{\sigma-}\rangle$ & 
           $\langle\theta_{\sigma-}\rangle$ & $m_t$ & $m_s$ \\ \hline
 SC$d$     &  $\frac{\pi}{2} I$ & $\frac{\pi}{2}I$ & $\frac{\pi}{2}I$ & $*$ &
              $+$  &     $+$    \\
 CDW + PDW &  $\frac{\pi}{2} I$ & $\frac{\pi}{2}(I+1)$ &
              $*$ & $\frac{\pi}{2}(I+1)$ &
              $-$  &     $+$    
\\
\hline\hline
\end{tabular}
\end{table}

From the numerical integration of the RG equations,
 we obtain the ground-state phase diagram shown in Fig.\ 1. 
The SC$d$ state (the CDW+PDW state) is obtained for 
  $ V_{\parallel}/U + V_{\perp}/U \gtrsim 0.4 ( \lesssim 0.4)$. 
The SC$d$ state, on the one hand, is stabilized by the on-site repulsive 
 interaction, which  
 segregates up-spin from  down-spin on the same site 
 and leads to the singlet pairing on a rung.  
On the other hand, the CDW+PDW state is obtained 
   due to the nearest-neighbor repulsive 
  interactions, which  induce  density wave leading 
   to the singlet state on the same site or chain.   
 The effect of $V_{\parallel}$ is slightly larger than that of $V_{\perp}$
 although both the intersite interactions have essentially the same effect of 
  inducing the CDW+PDW state. 
In Fig.\ 2, the change from the CDW+PDW state to the SC$d$ state is shown 
  with increasing the doping $\delta (> 0.05)$.
The novel aspect of the present paper is the competition induced by the  
 doping which reduces the effect of  only $V_{\parallel}$
  as shown in Eq.\ (\ref{eq:g}).
 In the inset, we show
  the respective masses estimated  
   from  $|m_a| \approx t_\parallel \, \exp(-l_a)$ ($a= t$, $s$, $\rho-$)
   by noting that  the corresponding coupling constant $|G_a|$ becomes 
   of the order of unity at $l = l_a$. 
Our system exhibits a \textit{second-order} phase transition 
  and
the magnon excitation gap vanishes at the quantum critical point (QCP).
The critical property for the Majorana triplet sector,
  which differs from that the conventional Tomonaga-Luttinger liquid,
   is described by the SU(2)$_2$ Wess-Zumino-Novikov-Witten model with 
  the central charge $c=3/2$.\cite{Gogolin_book} 
\begin{figure}[tb]
\begin{center}
\includegraphics[width=6.5cm]{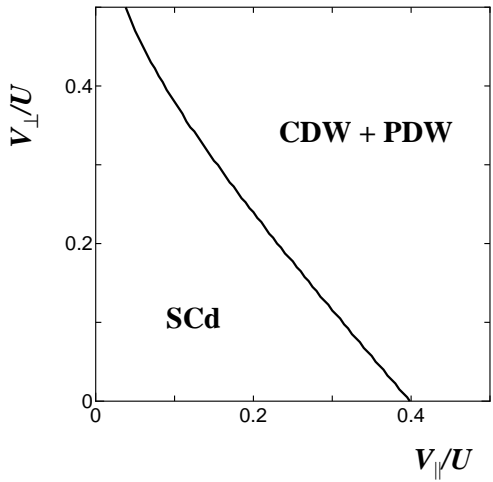}
 \vspace{-3mm}
\caption[]{
The ground-state phase diagram on the plane of $V_\parallel/U$ and
 $V_\perp/U$, with $U/t_\parallel=2$, $\delta=0.1$, and $t_\perp=t_\parallel$.
 }
\end{center}
\label{fig:phase}
\end{figure} 

\begin{figure}[tb]
\begin{center}
\includegraphics[width=6.5cm]{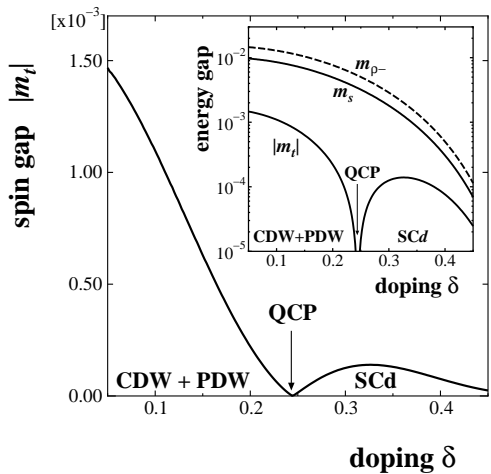}
\end{center}
\caption{
The doping dependence of the magnon spin gap $m_t$ with $U/t_\parallel=2$, 
 $V_\parallel/U=V_\perp/U=0.25$, and $t_\perp=t_\parallel=1$.
In the inset, the doping dependences of $m_{\rho-}$, $m_s$, and $m_t$ are
  are shown.
}
\label{gap}
\end{figure}

In the present paper, by applying the weak-coupling RG method to the
  EHM on two-leg ladder, we have shown that 
  the doping $\delta$ suppresses the CDW+PDW quasi-LRO state 
  and yields the system to the QCP, and that  the SC$d$
  quasi-LRO state is stabilized at further doping.
Here we discuss the experimental results of the two-leg ladder compound 
  Sr$_{14-x}$Ca$_x$Cu$_{24}$O$_{41}$.
The phase diagram of Sr$_{14-x}$Ca$_x$Cu$_{24}$O$_{41}$
   obtained in Ref.\ \citen{Vuletic} resembles 
   our phase diagram of  Fig.\ \ref{gap} 
 if the magnitude of gap $|m_t|$ is regarded as the transition temperature.
On closer look,
our phase diagram is contrast to the features of
  Sr$_{14-x}$Ca$_x$Cu$_{24}$O$_{41}$  that 
  the resistivity above the transition 
  temperatures  shows an insulating behavior and 
  there is no experimental evidence of the QCP 
    between the CDW state and the SC state.  
 In order to explain the phase diagram of Sr$_{14-x}$Ca$_x$Cu$_{24}$O$_{41}$,
  the dimensionality effect and/or the disorder effect
  has been discussed.\cite{Vuletic}
 The quantum critical behavior would be smeared out 
  by these effects, which   
  are not taken into account in the present paper.
However it will be still interesting to examine the competing region
 in the sense that  the magnon gap  would become extremely
  small and anomalous behavior can be expected 
   at temperatures higher than characteristic energies  of 
 the disorder and the dimensionality. 
We note that 
the origin of the high temperature insulating phase is still unknown and
 the analysis is left for a future study.

\acknowledgements
M.T.\ thanks A.\ Furusaki, H.\ Tsunetsugu, N.\ Fujiwara and H.\ Kitano 
for valuable discussions.
This work was supported by a Grant-in-Aid for 
Scientific Research on Priority Areas of Molecular Conductors 
(No. 15073213) from 
the Ministry of Education, Culture, Sports, Science and Technology,
   Japan.

\end{document}